# Coherent Lattice Vibrations in Superconductors

Alan M. Kadin*

Version June 2, 2007

Abstract:

A recent analysis by Kadin has noted that the superconducting wavefunction within the BCS theory may be represented in real-space as a spherical electronic orbital (on the scale of the coherence length $\xi_0$) coupled to a standing-wave lattice vibration. This lattice vibration, effectively a bound phonon, has wavevector $2k_F$ and a near-resonant frequency (on the order of the Debye frequency) that maximizes the attractive electrostatic interaction energy with the electronic orbital. The present paper extends this picture to a coherent standing-wave pattern of electron and phonon waves that traverses the entire superconductor on the macroscopic scale. These parallel planes form a diffractive waveguide for electron waves traveling parallel to the planes, permitting lossless supercurrent. A similar picture may be extended to unconventional superconductors such as the cuprates, with an array of standing spin waves rather than phonons. Such coherent lattice vibrations should be universal and distinctive indicators of the superconducting state, and should be observable below $T_c$ using standard x-ray and neutron diffraction techniques. Further implications of this picture are discussed.

*Address: Princeton Junction, NJ 08550, USA

Email: amkadin@alumni.princeton.edu

I. Introduction

The BCS theory of superconductivity is well established, with long-range order associated with correlations of quantum waves for Cooper pairs. The Cooper pair, in turn, is based on a net electron attraction mediated by the electron-phonon interaction. This theory has been highly instructive for conventional low-temperature superconductors, and has served as a model for other theories for unconventional superconductors such as the cuprates. However, the conventional theory does not provide a clear real-space physical picture that ties together microscopic interactions with long-range superconducting order.

The present paper provides such a picture, in which standing waves of electrons and phonons (with $k = 2k_F$) extend coherently across a superconducting sample. Although electron standing waves are associated with insulating behavior in the classic band theory of crystals, it is shown here that standing wave planes may form a diffractive waveguide for traveling waves parallel to the planes, which represent lossless supercurrents. Futhermore, such a picture is also properly compatible with macroscopic fluxoid quantization $\Phi = n\Phi_0 = nh/2e$, despite the absence of an explicit quantum wavefunction associated with electron pairs. While this picture seems to be sharply at odds with the conventional understanding of superconductivity, it appears to be compatible with the standard BCS and Ginzburg-Landau theories.

The focus is primarily on conventional phonon-mediated superconductivity, but this picture is directly extendable to other excitations and other symmetries. For example, if electron standing waves are out-of-phase for the two spin directions, this will permit them to couple to spin waves (magnons). This may be relevant to unconventional superconductors such as the cuprates.

If the phonon standing wave patterns are indeed real, they should be directly observable by x-ray and neutron diffraction. The diffraction lines may be much weaker than those from crystalline planes, since the corresponding superconducting energy gap is much smaller (by ~ 1000) than the energy gap of a typical insulator. However, comparison of diffraction spectra above and below the superconducting transition (induced by either temperature or magnetic field) should permit identification of planes associated with superconducting order.

This picture is based on an electron orbital picture of Kadin [1], which in turn is based on the BCS equations. This microscopic orbital picture is derived in Section II, and its extension to macroscopic coherence is described in Section III. Section IV shows how these macroscopic wave patterns give rise to the usual supercurrent and flux quantization. Section V describes how this entire picture can be extended to alternative mechanisms for superconductivity, for example involving macroscopic spin waves, and other issues that may be relevant to the high-temperature superconducting cuprates. Guidelines for observing the predicted macroscopic coherent lattice and magnetic oscillations are discussed in Section VI.

II. Orbital Picture

The BCS theory is typically analyzed in k-space. To the extent any real-space picture is used to motivate these equations, it usually involves two point electrons ($+k\uparrow$ and $-k\downarrow$) moving in opposite directions through a lattice of positive ions [2]. The positive ions distort towards the electrons, leaving a positive trail behind each moving electron. Each electron is attracted to the positive trail of the other, creating a Cooper pair. However, electrons in solids are not really point particles on the atomic scale, and this picture seems unsatisfying as a bound quantum state.

An alternative real-space picture has also been presented [1, 2], but much more rarely. In this picture, the two electrons form a spherical quasi-atomic S-wave orbital on the scale of the coherence length $\xi_0$, composed of outgoing and incoming waves for k near $k_F$, yielding spherical standing waves with nodes separated by a distance $\pi/k_F$ (see Fig. 1). Indeed, one obtains precisely such a wavefunction if one Fourier transforms the pair wavefunction from the BCS theory. (The mean envelope of the wavefunction turns out to be $\pi\xi_0$ [2].) The connection between these two pictures can be observed if one notes that this wavefunction has a charge distribution ($\sim|\Psi|^2$) with spatial modulation at $2k_F$. Such a charge modulation will attract the positive lattice ions, creating an induced matching modulation in the lattice. The electrostatic interaction with this lattice charge modulation permits electrons of both spins to align their charge densities, which would otherwise be energetically unfavorable.

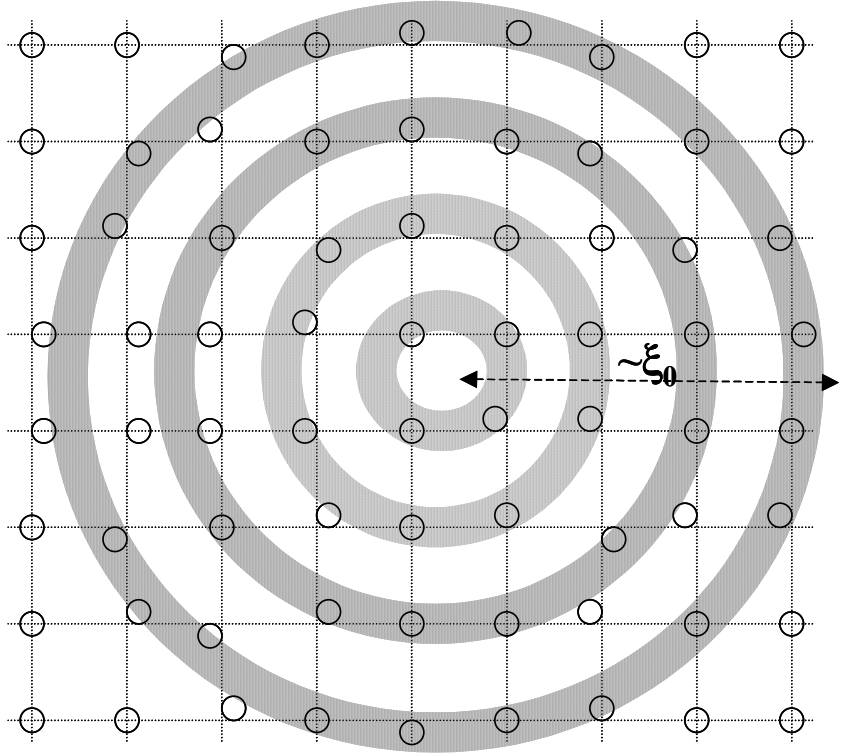

*Fig. 1. Conceptual picture of the charge density associated with an electron orbital that is part of the superconducting ground state, together with lattice distortions that couple to it. The antinodes of the electronic orbital (shaded) attract the positive lattice ions. More generally, this is a dynamic interaction associated with phonon standing waves.*

The interaction will be even stronger if the lattice is driven close to a resonant frequency on the order of the Debye frequency $\omega_D$, corresponding to standing-wave phonons with $k = 2k_F$. In this case, the lattice charge distribution oscillates at $\sim\omega_D$, and the electronic charge distribution adiabatically adjusts to remain in phase with the phonons. The resonantly driven charge density in the lattice may be even larger than the driving charge density in the electrons ("overshoot"), thus overcoming Coulomb repulsion and corresponding to a bound quantum state of both electrons. This two-electron orbital may be regarded as the physical realization of a Cooper pair, although as discussed below, it is really only a small piece of a many-electron state.

The picture of Fig. 1 may also be examined from the point of view of wave diffraction. Unlike the case of an electron orbital in an atom, there is no strong central force confining the electron waves. Instead, the outgoing spherical wave at $k_F$ is diffracting off the induced lattice modulation at $2k_F$, becoming an incoming spherical wave at $-k_F$. As the phonon oscillates, the phase of the diffracted wave also changes to maintain the proper phase relations. Furthermore, one can think of the phonon standing wave as consisting of outgoing and incoming components which are also diffracted from the electronic charge modulations. This provides a real-space wave-based illustration of an electron-phonon interaction, which is believed to provide the physical basis for superconductivity in conventional low-temperature superconductors. But note that these phonons are not randomly phased thermal fluctuations; they are synchronized across the orbital, and are part of the superconducting ground state.

Although this orbital picture is described here in terms of phonons, it should be clear that it could apply equally well to other wave modes and other orbital symmetries. This will be discussed further in Section V.

III. Long-Range Coherence

The picture above of a single Cooper pair is neither complete nor fully consistent. First, within the range $\xi_0$, there are many overlapping Cooper pairs, typically $\sim 10^6$. The charge modulations for each of these will interfere randomly with the others, unless they are properly synchronized. Second, all of the different domains on the $\xi_0$ scale also need to be synchronized with all of the others, if one is to have long-range phase coherence associated with macroscopic supercurrents.

One can see the meaning of wave coherence by reconsidering the classic Huygens principle for propagating waves (Fig. 2). One can construct a propagating wave by regarding each point on the wavefront as a source of outgoing spherical waves. Equivalently, it is well known that the phase delays among a phased array of isotropic sources (such as radio transmitters) can be adjusted to tune the resulting coherent beam in any direction. The macroscopic coherent wave pattern consists (at least locally) of a set of parallel planes of equal phase. In the present case, one has an array of point sources of standing spherical waves, with both outgoing and incoming components having a range $\sim \xi_0$, but Huygen's principle should be equally applicable as long as the sources are strongly overlapping. So one would expect a macroscopic array of parallel nodal planes, consisting of both electron and phonon waves as in the orbital sources.

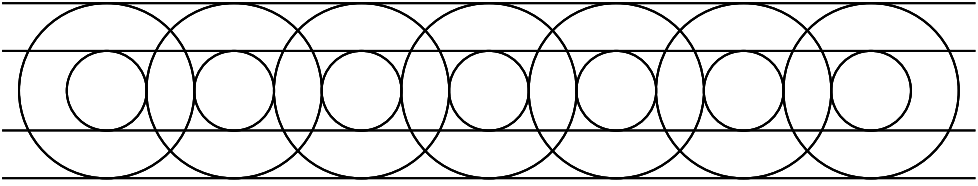

*Fig. 2. Picture representing the construction of a coherent one-dimensional wave by superposition of many local spherical orbitals such as that in Fig. 1, similar to Huygens Principle.*

A simple argument shows why coherence should occur among N overlapping local sources. If each source is incoherent with the others, then the mean phonon field from each of the N-1 other sources will just provide random noise that will not contribute to the binding energy of a given orbital. In contrast, if all N sources are in phase, then the total induced phonon wave will be a factor of N larger than that due to a single source, and the interaction energy per orbital will be a factor of N larger than the incoherent case. The energy gap $2\Delta$ and the coherence length $\xi_0 \sim 1/\Delta$ correspond to this enhanced coherent condition.

This coherence argument also provides an intuitive picture of nucleation of superconductivity from the normal state. One can imagine that a single electron near the Fermi surface scatters from a fluctuating phonon with $2k_F$, creating a local electron standing wave, which generates charge density modulation that in turn reinforces the phonon. This growing phonon modulation attracts other electrons to be spatially synchronized, creating an instability with positive feedback until all of the available electrons in the region have been included. Different regions may nucleate into different nodal patterns (domains), but the mismatches at the domain surfaces would be energetically unfavorable, leading eventually to full long-range coherence.

The nodal pattern is not generally unique, and reflects symmetry-breaking from an initial isotropic condition. The actual pattern may be determined by a combination of factors, including boundary conditions, currents and fields, local defects, and crystalline anisotropy. It is argued below that the nodal planes will tend to orient so that supercurrent flows parallel, rather than perpendicular, to the planes.

It is important to appreciate that the electron quantum wavefunctions themselves are *not* being synchronized. That would violate the Pauli exclusion principle, since the electrons are fermions. It is only the charge densities associated with electron standing waves that are in phase (in space and time). The electron quantum waves themselves may have a range of ω and k similar to that in the normal state.

Analogous spatial coherence of electron standing waves is well known to occur at the band edge in the classic band theory of crystals. In that case, the charge modulation in the lattice is fixed, and the energy gap is associated with standing waves produced by electron diffraction from this fixed lattice. Standing waves that are in-phase with the lattice modulation lie below the gap; those out-of-phase lie above the gap (Fig. 3). The present picture of superconductivity is similar, except that the diffracting lattice is

1) One-dimensional rather than three-dimensional,
2) Induced, rather than pre-determined,
3) Incommensurate with the crystal lattice, and
4) Dynamic rather than static.

There are also some similarities with charge-density waves (CDW) that tend to occur in quasi-one-dimensional crystals [3-5]. A CDW is an static instability in a one-dimensional lattice below some critical temperature, in which an energy gap forms at the Fermi surface, turning a conductor into an insulator. (The CDW gap equation is similar to the BCS gap equation [3].) This is associated with coherent electron standing waves coupled to a lattice distortion at $2k_F$, sometimes called a "frozen phonon", which is typically commensurate with the crystal lattice. The coherent waves in the present picture of superconductivity might be regarded as dynamic, incommensurate CDWs. However, these properties permit the transport of lossless currents in the superconductor, as described below in Section IV.

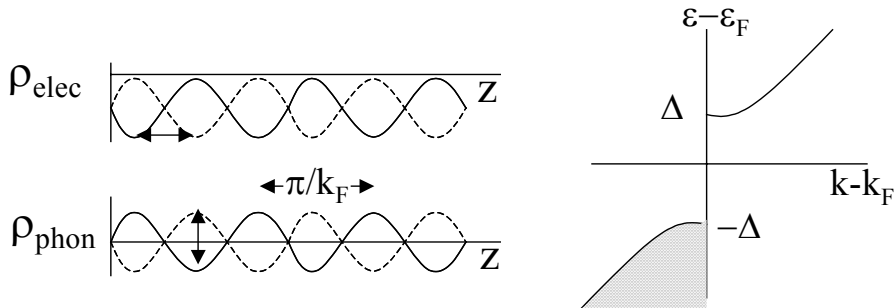

*Fig. 3. The standing-wave charge modulation patterns of the electrons and the lattice give rise to the superconducting energy gap, similar to the band theory of crystals. The electronic charge distribution drives the phonon modulation, which in turn diffracts the electron waves.*

IV. Supercurrents and Flux Quantization

Superconductivity is generally presented as a macroscopic quantum state, and indeed, supercurrents, the Meissner effect, and flux quantization can all be easily derived by assuming (as in the London and Ginzburg-Landau equations) a macroscopic quantum wavefunction $\Psi = |\Psi|\exp(i\phi)$, where $\hbar\nabla\phi = 2mv_s - 2eA$, corresponding to charge carriers of mass 2m and charge 2e. Furthermore, a Cooper pair of charge 2e would be a boson rather than a fermion, so one could in principle have all such pairs in the same quantum wavefunction.

However, neither distinct Cooper pairs nor such a quantum wavefunction are evident in the dynamic standing-wave picture described above. Despite that, we show below that this picture leads to macroscopic traveling waves, guided parallel to the standing wave planes, that support supercurrents and yield flux quantization with $\Phi_0 = h/2e$.

Consider first the case where a weak magnetic field is applied parallel to the surface of a superconductor (Fig. 4). Then the vector potential A (in an appropriate gauge) is parallel to the surface, and assume that the nodal planes are also parallel to the surface. For all electron states, $\hbar k = mv - eA$, so if the entire Fermi sea shifts by $mv_s = eA$ in velocity space, one has currents given by the usual London equation $J_s = n_s e v_s = \Lambda A$, where $\Lambda = m/n_s e^2$. Since the quantum phases of the electronic states are unchanged, so are the phases of the macroscopic standing waves; one has $\nabla\phi=0$ or $\phi$ = constant. This also corresponds to the Meissner effect with screening currents parallel to the surface up to a magnetic penetration depth $\lambda_L = \sqrt{(\Lambda/\mu_0)}$. These screening currents are lossless not because this macroscopic wave is a quantum wave function, but rather because there is an energy gap, with no available states nearby. It seems remarkable that lossless currents can flow due to the same mechanism (electron standing waves) that makes other materials insulating, but that seems to be a consistent interpretation.

But then why do band insulators not exhibit diamagnetism? A 3-D insulator exhibits standing waves in all directions, which may block such screening currents. On the other hand, some measurements of 1-D CDW insulators have indicated evidence of lossless diamagnetism [5].

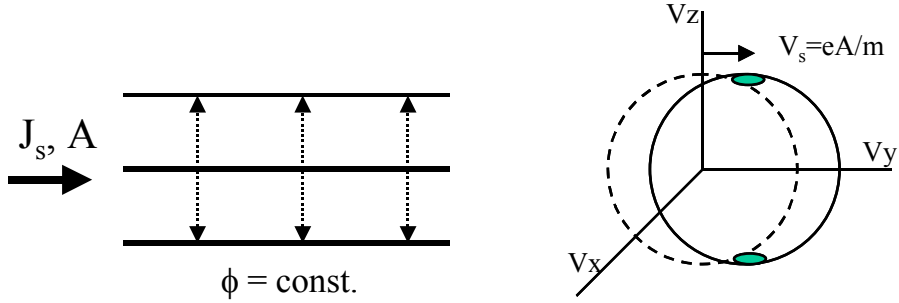

Fig. 4. *Diamagnetism in the Meissner effect is associated with supercurrent flow parallel to the nodal planes and a shift in the Fermi sphere, but with the phase pattern unchanged.*

Of course, the primary hallmark of superconductivity is transport supercurrent, driven by a phase gradient, rather than just diamagnetism. This requires not merely $J_s > 0$, but also a traveling wave with $k > 0$. Consider a superconducting wire where the nodal planes form concentric cylinders parallel to the surface (Fig. 5). This can form the basis for waveguide modes of wave propagation, that combine a traveling wave along the axis with standing waves perpendicular to the axis. Such a waveguide mode can be constructed from waves with wavelength $\lambda_0$, tilted at an angle $\theta$ toward the axis. Such modes are well known for electromagnetic waveguides [6], and are characterized by an enhanced wavelength $\lambda = \lambda_0/\sin\theta$, and an equally enhanced phase velocity $v_{ph} = \lambda f$ which can even be greater than the speed of light. The group velocity, of course, is slow, given all of the zig-zag motion of the constituent waves.

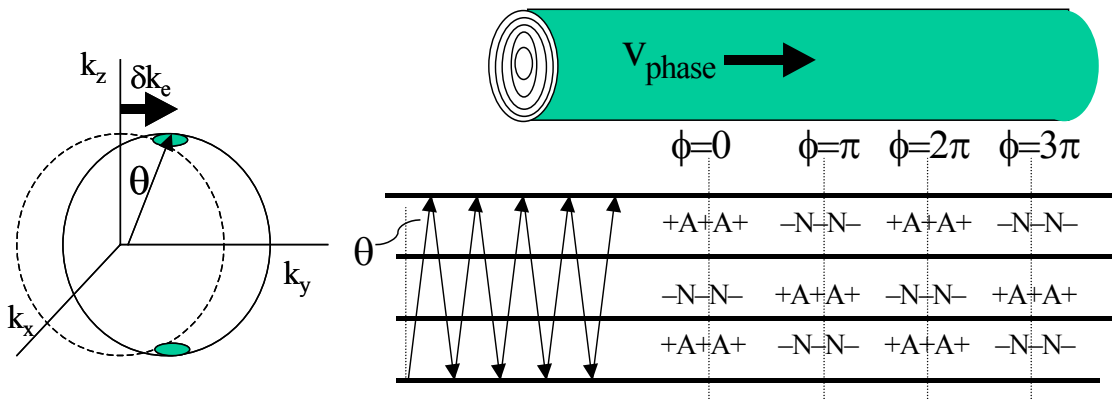

Fig. 5. *For transport current down a superconducting wire, the standing wave is modified slightly to contribute a phase gradient and traveling wave component, with a phonon (max/min = +/- ) and electron charge wave (max/min = Antinode A/Node N).*

For the superconductor case, the constituent waves are coupled charge density waves and phonons, both with $k = 2k_F$, so that $\lambda_0 = \pi/k_F$. The entire Fermi sphere shifts in k-space by $\delta k_e = (mv_s - eA)/\hbar$, corresponding to a small angle $\theta = \tan^{-1}(\delta k_e/k_F)$ relative to the standing waves. Near the surface of the wire, $\delta k_e = mv_s/\hbar$, so that $\theta \approx v_s/v_F$ and $\lambda = h/2mv_s$, the same formula as a quantum wave for a particle with mass 2m. However, this is not a simple quantum wave of a pair of electrons (or a superposition thereof); in the z-direction it is a high-order standing wave. In the center of the wire (beyond $\lambda_L$ from the surface), $\theta \approx eA/\hbar k_F$, so $\lambda \approx h/2eA$, which is the form for a quantum wave of a particle of charge 2e. Overall, this yields a macroscopic guided wave with phase gradient $\nabla\phi = k_c = (2mv_s - 2eA)/\hbar$, consistent with Ginzburg-Landau theory for a pair wavefunction. This corresponds to a uniform phase through the cross-section of the wire, and a uniform phase gradient along a wire of uniform cross-section.

If one connects this wire into a loop, the phase difference in going around the wire must be a multiple of $2\pi$, which yields in the usual way that $\int A \cdot dl = \Phi = n\Phi_0 = nh/2e$. Similarly, a vortex consists of concentric cylindrical shells of standing-wave planes, with a circumferential phase difference of $2\pi$, corresponding to a single flux quantum. Further, the angular phase velocity of the wave is $\omega$, the phonon frequency, independent of the radius r. This means that $v_{ph} \propto r$, while $v_s \propto 1/r$. The vortex core corresponds to a region where $\Delta\phi > \pi$ within a distance $\sim \xi_0$. This weakens the interference effect, reducing the strength of the coherent modulation and hence the energy gap in the vortex core.

V. Extensions to HTS

While the specific terminology for the coupled standing waves above referred to lattice distortions and conventional phonon-mediated superconductivity, the picture itself is not so restrictive. Indeed, any polarizable mode that couples to electronic charge modulation at $2k_F$ could contribute to the superconducting energy gap. That might include, for example, electronic coupling to filled shells or other sublattices, or to excitons or plasmons. If there are multiple alternative modes, one would expect that the electron standing waves will preferably couple to whichever mode provides the largest energy gap.

In a dynamic charge density wave as we have discussed, the electrons with opposite spins will oscillate in phase, maximizing the modulation in the charge density, but removing any modulation in the spin density. Alternatively, one might have a dynamic spin density wave, in which the electrons with opposite spins oscillate out of phase, which would maximize the modulation in the electronic spin density but cancel the charge density modulation [1]. This would be relevant to a material that has magnetic excitations. In the diffraction language, a polarized electron wave diffracts from a standing spin wave, which mutually reinforce one another to yield a macroscopically coherent wave with an energy gap. The patterns would be much the same as for the phonon case, but the oscillation frequency would be characteristic of spin waves in the material with $k = 2k_F$. This might correspond to some models of superconductivity in the high-$T_c$ cuprates, for example.

A further generalization of the picture includes the possibility of alternative symmetries of the underlying orbitals. In particular, the standard BCS theory assumes S-wave symmetry, but P-wave or D-wave symmetries may be energetically favored in some cases. Then the orbitals that act as the point sources of the coupled standing waves would each have lobes and angular nodes pointing in preferred directions. The nodal directions, which would refer to specific directions in the crystal lattice, would be maintained in the macroscopic waves created by coherent superposition of these local sources. For example, the cuprates are generally believed to exhibit $d_{x^2-y^2}$ symmetry, with lobes directed in the a- and b-directions of the a-b crystalline plane [7]. Superimposing many such orbitals would lead to conducting channels (and parallel standing-wave planes) oriented along the a- or b-axis, but not halfway between these axes.

The analysis earlier in this paper assumed an isotropic material with a spherical Fermi surface. In contrast, the cuprates are strongly anisotropic, consisting of conducting a-b planes weakly coupled in the c-axis direction. If one focuses on a single a-b plane, one can envision parallel standing waves within the plane, with conducting channels along parallel stripes, along either the a- or b-direction. Adding the third dimension transforms the stripes into parallel planes, perpendicular to the a-b planes. These standing-wave stripes or planes could re-orient, depending on the directions of the current in the planes.

VI. Measuring Coherent Lattice Oscillations

The central feature of this diffractive-wave picture of superconductivity is the presence of macroscopically coherent oscillating gratings at $2k_F$. These should be measurable using structural analysis techniques such as x-ray and neutron diffraction. But why haven't they been seen already? After all, this is a region of k-space that has been well explored for many years.

However, there are several aspects of these standing waves that may make them relatively difficult to observe. First, they should be much weaker than typical crystalline diffraction peaks. Electron diffraction of crystal lattices yields an energy gap ~ 1 eV, while similar diffraction from the induced phonon lattice yields an energy gap ~ 1 meV. If x-ray or neutron diffraction amplitudes from the induced lattice are weaker by the same factor, that would suggest a diffracted power (which goes as the square of the diffracted amplitude) that might be up to six orders of magnitude weaker than standard crystalline diffraction peaks. A signal on this scale could easily be overlooked, particularly in a spectral region where crystal structure from impurity phases may be signficant, or it might be below the noise level of the measurement system. Second, these structures can only be observed below the critical temperature $T_c$, and further would appear only gradually below $T_c$, since superconductivity is a second-order phase transition. This suggests that a successful strategy for detection of these structures would involve switching the same sample (perhaps repeatedly) between the superconducting and the normal states, either with temperature or possibly with magnetic field, and looking carefully for small changes. Finally, the predicted gratings are dynamic rather than the usual static gratings associated with crystal structures. That means that the diffracted phase oscillates at the phonon frequency (10-100 THz). If the intrinsic linewidth of the detector is narrower than this, the detector would integrate over multiple periods, and the measured diffraction would be sharply attenuated. This is probably not relevant for keV x-ray detectors, but might be significant for electron diffraction.

An additional consideration, for magnetic excitations that may be associated with cuprate superconductors, is to use waves that couple magnetically. The standard technique uses polarized neutron beams, since neutrons have a spin (magnetic moment) despite the lack of electric charge. Alternatively, as electromagnetic waves, x-rays also interact magnetically, and x-ray magnetic diffraction may also be used for detecting these structures at $2k_F$.

It is also possible that relevant structures have already been observed in some cases, but their general significance may not have been fully appreciated. For example, there have been various observations in recent years of static and dynamic stripes in different cuprate superconductors [8,9], which some have considered important for understanding the mechanism of superconductivity in these materials. It remains to be seen whether some of these observations may be relevant to the general picture presented here.

Another key aspect of these patterns is their dynamic nature. The picture as presented suggests that the oscillations should be close to a resonance for $2k_F$, but does not specify the frequency. Further, the picture suggests that a gap should open up in the phonon (or magnon) excitation spectrum due to diffraction from the electrons, similar to the superconducting energy gap. The phonons phased properly should lie just below this energy (frequency) gap and be strongly occupied, while those out of phase would lie just above the gap and be mostly empty, at least for $T \ll T_c$. Inelastic neutron scattering or inelastic x-ray scattering are often used to measure phonon spectra; other techniques that are sometimes used include optical Raman scattering, far-infrared reflectance, and electronic probes such as inelastic tunneling spectroscopy. For magnetic excitations, inelastic scattering of polarized neutrons may be used. For either type of excitations, a set of careful measurements of dispersion relations and density of states, both above and below $T_c$, could more clearly establish the validity of this approach.

VII. Conclusions

In summary, the present paper suggests a fundamental reconsideration of the physical basis for superconductivity, even for conventional metals where a phonon mechanism has long been established. While the equations of the BCS theory are undoubtedly valid, their interpretation in terms of a macroscopic quantum wavefunction of Cooper pairs may be questionable. In the alternative real-space diffracting-wave picture, superconductors incorporate a universal coherent array of one-dimensional diffracting planes with $k = 2k_F$, dynamic and incommensurate with the crystal structure. Electron diffraction from these planes yields the superconducting energy gap $2\Delta$ in direct analogy to the energy gap at the band edge in crystals. For conventional low-temperature superconductors, the planes are phonon standing-waves at $\omega \sim \omega_D$, coupled to dynamic electron charge density waves. For high-temperature superconducting cuprates, the planes may be spin standing-waves, coupled to dynamic electron spin density waves. Further, these planes should form a macroscopic coherent waveguide for supercurrents parallel to the planes, across the entire length of the superconductor. These supercurrents are associated with a macroscopic traveling wave that generates flux quantization and vortices, including the factor "2e" that is usually associated with Cooper pairs. However, the component electrons of this coherent wave are not strictly paired and do not have the same energy; rather, their charge (or spin) distributions are coherently aligned with a macroscopic phonon (or magnon) wave.

Of course, such a novel picture requires experimental evidence for it to be taken seriously. The diffracting planes should be measurable with x-ray or neutron diffraction below $T_c$, and the dynamic oscillations should be evident in modified phonon (or magnon) excitation spectra in inelastic scattering. If these are eventually observed, this diffractive-wave picture of superconductors may contribute to a profound shift in our understanding of the nature of coherent states in many-body systems.